\documentclass[two column]{jpsj2}

\title{Fermi Arc in Doped High-T$_c$ Cuprates
}

\author{Takashi Yanagisawa$^{1}$, Mitake Miyazaki$^{1,2}$ and Kunihiko Yamaji$^1$}

\inst{$^1$Condensed-Matter Physics Group, Nanoelectronics Research Institute,
National Institute of Advanced Industrial Science and Technology, Tsukuba Central 2,
1-1-1 Umezono, Tsukuba 305-8568, Japan\\
$^2$Department of Physics, Aoyama Gakuin University, 5-10-1 Fuchinobe, Sagamihara,
Kanagawa 229-0558, Japan\\
}

\recdate{\today}

\abst{
We propose a $d$-density wave induced by the spin-orbit coupling in the
CuO plane.
The spectral function of high-temperature superconductors
in the under doped and lightly doped regions
is calculated in order to explain the Fermi arc spectra observed recently by
angle-resolved photoemission spectroscopy.
We take into account the tilting of CuO octahedra as well as the on-site Coulomb
repulsive interaction; the tilted octahedra induce the staggered
transfer integral between $p_{x,y}$ orbitals and Cu $t_{2g}$ orbitals, and
bring about nontrivial effects of spin-orbit coupling
for the $d$ electrons in the CuO plane.
The spectral weight shows a peak at around ($\pi/2$,$\pi/2$) for light doping 
and extends around this point forming an arc as the carrier density increases, 
where the spectra for light doping grow
continuously to be the spectra in the optimally doped region.
This behavior significantly agrees with that of the angle-resolved photoemission
spectroscopy spectra.
Furthermore, the spin-orbit term and staggered transfer effectively induce a 
flux state, a pseudo-gap with time-reversal symmetry breaking.
We have a nodal metallic state in the light-doping case since the pseudogap has
a $d_{x^2-y^2}$ symmetry.
}

\kword{spectral function, high-Tc cuprate, distortion, spin-orbit coupling, stripes}

\begin{document}
\sloppy
\maketitle


In recent years, oxide high-$T_c$ superconductors have been investigated
intensively; 
anomalous metallic behaviors as well as high critical temperatures $T_c$
have been focused in the study of high-$T_c$ cuprates.
Clarifying the origin of an anomalous metal with a pseudogap is also a
challenging problem
attracting many physicists.
Recently, a peak crossing the Fermi level in the node direction of the $d$-wave
gap has been observed in a lightly doped La$_{2-x}$Sr$_x$CuO$_4$ (LSCO) by 
angle-resolved photoemission
spectroscopy (ARPES).
The spectral weight shows a peak at around $(\pi/2,\pi/2)$ for the light-doping
case and it extends along the Fermi surface with increasing carrier
density.[1,2]
The existence of incommensurate correlations has also been reported by 
neutron-scattering measurements,
suggesting vertical stripes in an underdoped region
and diagonal stripes in a lightly doped region.
Modulation vectors are $Q_s=(\pi\pm 2\pi\delta,\pi)$,
$Q_c=(\pm 4\pi\delta,0)$ (or $Q_s=(\pi,\pi\pm 2\pi\delta)$,
$Q_c=(0,\pm 4\pi\delta)$) for the hole-doping rate $x\ge 0.05$, where $\delta$
indicates the approximately linear dependence 
$\delta=x$.[3-5]

In the lightly doped region for $x< 0.05$, the suggested modulation vector is
$Q_s=(\pi\pm 2\pi\delta,\pi\pm 2\pi\delta)$ and the deviation from the linear
dependence exhibits  $\delta<x$.[4]
From the experiments of resistivity in the lightly-doped region of LSCO, 
the system
holds a metallic behavior below $T_N$, which may be due to the formation of 
metallic
charge stripes.[6]

The purpose of this paper is to investigate the spectral function of doped 
high-temperature
superconductor in order to elucidate the recently observed ARPES
Fermi arc spectra, taking into account the stripe orders suggested by 
neutron-scattering measurements.
We emphasize the importance of distortions of CuO octahedra.

The tilting of CuO octahedra induces transfer integrals between $p_{x,y}$ 
orbitals
and Cu $t_{2g}$ orbitals, and then the spin-orbit coupling between $t_{2g}$ 
orbitals is not decoupled from $e_g$ networks.[8,9]
For the spin-orbit coupling 
\begin{equation}
H_{so}=\xi{\bf \ell}\cdot{\bf s}, 
\end{equation}
matrix elements exist between $t_{2g}$ orbitals:
\begin{equation}
\langle d_{xz}({\bf r})\uparrow|H_{so}|d_{yz}({\bf r})\uparrow\rangle=-(i/2)\xi,
\end{equation}
\begin{equation}
\langle d_{yz}({\bf r})\uparrow|H_{so}|d_{xz}({\bf r})\uparrow\rangle=(i/2)\xi,
\end{equation}
\begin{equation}
\langle d_{xz}({\bf r})\downarrow|H_{so}|d_{yz}({\bf r})\downarrow\rangle=(i/2)\xi,
\end{equation}
\begin{equation}
\langle d_{yz}({\bf r})\downarrow|H_{so}|d_{xz}({\bf r})\downarrow\rangle=-(i/2)\xi.
\end{equation}
We have also matrix elements between $t_{2g}$ orbitals and $e_g$ orbitals:
\begin{equation}
\langle d_{xz}({\bf r})\downarrow|H_{so}|d_{x^2-y^2}({\bf r})\uparrow\rangle=\xi/2,
\end{equation}
\begin{equation}
\langle d_{yz}({\bf r})\downarrow|H_{so}|d_{x^2-y^2}({\bf r})\uparrow\rangle=-i\xi/2,
\end{equation}
and those for reversed spins obtained by multiplying by $-1$.
The matrix elements induced by the tilting are
\begin{equation}
\langle p_x(x-a/2,y)\sigma|H_{pd}|d_{xz}({\bf r})\sigma\rangle=-t_{xz}{\rm e}^{i{\bf Q}\cdot{\bf r}},
\end{equation}
\begin{equation}
\langle p_y(x,y-a/2)\sigma|H_{pd}|d_{yz}({\bf r})\sigma\rangle=-t_{yz}{\rm e}^{i{\bf Q}\cdot{\bf r}},
\end{equation}
where ${\bf r}=(x,y)$, ${\bf Q}=(\pi,\pi)$, $a$ is the lattice constant 
and $H_{pd}$
denotes the hybridization term.
The factor ${\rm e}^{i{\bf Q}\cdot{\bf r}}$ induced by the tilting of octahedra
in a staggered manner, which leads to the doubled unit cell.
$t_{xz}$ and $t_{yz}$ are assumed to be given by sin$\theta$ for the tilt 
angle $\theta$.
In the low-temperature tetragonal (LTT) phase, the spin-orbit coupling may be
significantly smaller than that
in the low-temperature orthorhombic (LTO) phase, since
the integrals between oxygen $p$ and Cu $t_{2g}$ orbitals remain zero
along the tilt axis, where oxygen atoms never move: $t_{yz}=0$ if the tilt 
axis is in the y direction.

\begin{figure}[t]
\includegraphics[width=\columnwidth]{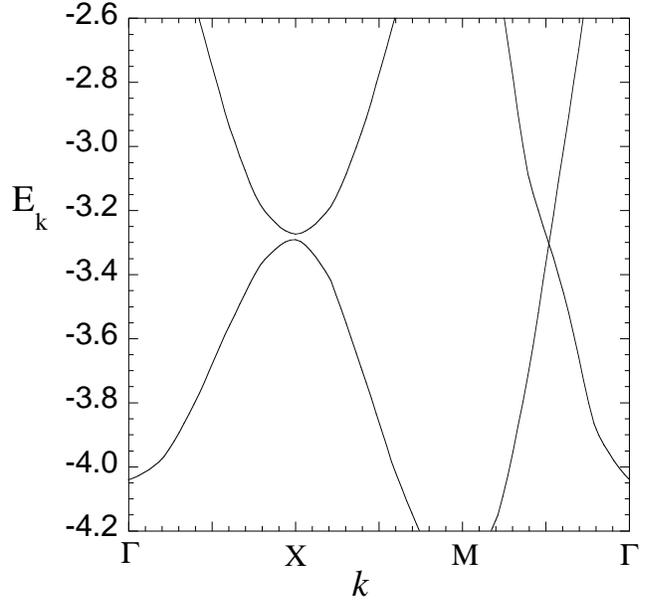}
\caption{
Dispersion relation for the three-band $d$-$p$ model.
The parameters are as follows: $\epsilon_{d_{x^2-y^2}}=-2$, $\epsilon_p=2$, 
$\epsilon_{d_{xz}}=\epsilon_{d_{yz}}=-1$, $t_{pp}=0.2$,
$\xi=0.1$, and $t_{xz}=t_{yz}=0.3$ in units of $t_{pd}$, where $t_{pd}$ is the transfer
between $d$ and $p$ orbitals and $t_{pp}$ is that between neighboring oxygen $p$
orbitals.  The splitting at $X=(\pi,0)$ is of the order of 10meV.  Each curve is
a twofold degeneracy, i.e., Kramers degeneracy. 
}
\label{ek}
\end{figure}

\begin{figure}[t]
\includegraphics[width=\columnwidth]{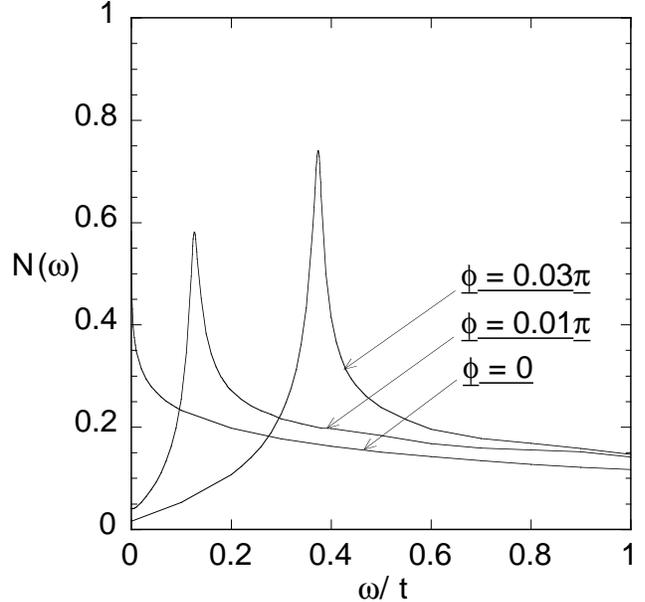}
\caption{
Density of states for $\phi=0$, 0.01$\pi$ and $0.03\pi$.
We set $t'=0$.
}
\label{dos}
\end{figure}

The dispersion in the presence of distortions for the five-band $p$-$d$ model is
shown in Fig.\ref{ek}.
The parameters are as follows: $\epsilon_{d_{x^2-y^2}}=-2$, $\epsilon_p=0$, 
$\epsilon_{d_{xz}}=\epsilon_{d_{yz}}=-1$, $t_{pp}=0.2$,
$\xi=0.1$, and $t_{xz}=t_{yz}=0.3$ in units of $t_{pd}\sim 1eV$ where $t_{pd}$ 
is the transfer
between $d$ and $p$ orbitals and $t_{pp}$ is that between neighboring oxygen $p$
orbitals.  For the parameters shown above, the splitting at $X=(\pi,0)$ is of 
the 
order of 10meV$\approx$100K.  Each curve is a twofold degeneracy, i.e. Kramers 
degeneracy as it should be. 
This structure near the Fermi energy is well understood, using the single-band 
model with the reduced Brillouin zone, as
\begin{equation}
H_0= \sum_{k\sigma}[ \xi_k c^{\dag}_{k\sigma}c_{k\sigma}
+\Delta_{{\bf k}\sigma} c^{\dag}_{k\sigma}c_{k+Q\sigma} ],
\label{hamil}
\end{equation}
where
\begin{equation}
\xi_k=-2t({\rm cos}(k_x)+{\rm cos}(k_y))-4t'{\rm cos}(k_x){\rm cos}(k_y), 
\end{equation}
and $Q$ denotes the wave vector $Q=(\pi,\pi)$.
$\Delta_{{\bf k}\sigma}$ is a complex parameter satisfying
\begin{equation}
\Delta_{{\bf k}+Q\sigma}=\Delta_{{\bf k}\sigma}^*.
\end{equation}
$\Delta_{{\bf k}\sigma}$ is taken as[10,11]
\begin{equation}
\Delta_{{\bf k}\sigma}=i\sigma{\rm sin}\phi\cdot(-2t)Y({\bf k})
\end{equation}
with
$Y({\bf k})={\rm cos}(k_x)-{\rm cos}(k_y)$ in order to reproduce the dispersion for
the five-band CuO model near the Fermi energy, where $\phi=\lambda\theta$;
$\theta$ is the tilt angle between the Cu-O plane and the Cu-O bond,
and $\lambda$ is a constant estimated as
$\lambda\approx 0.2$.[11]
The dispersion relation of the noninteracting part $H_0$ in eq. (\ref{hamil}) is

\begin{equation}
E_{{\bf k}}=\frac{1}{2}[\xi_{{\bf k}}+\xi_{{\bf k}+Q}
-\sqrt{(\xi_{{\bf k}}-\xi_{{\bf k}+Q})^2+4|\Delta_{{\bf k}\sigma}|^2}].
\end{equation}
At half-filling, the low-energy excitations are described by the Dirac 
fermion since
there is a Fermi point at $(\pm\pi/2,\pm\pi/2)$.  A light hole doping 
results in a
small Fermi surface around this point with the excitation gap near $(\pi,0)$.
This peculiar feature induces a pseudogap in the density of states and we obtain
a nodal metallic state.
The pseudogap induced by the spin-orbit coupling has a $d_{x^2-y^2}$ symmetry as
apparent from that of $\Delta_{{\bf k}\sigma}$.
The density of states for small $\phi$ shown in Fig. \ref{dos} clearly indicates
a pseudogap structure for a small excitation energy.

The eigenstate of $H_0$ resembles the $d$-density wave proposed for an anomalous
metallic state in the underdoped high-$T_c$ cuprates[12],
and possesses the following order parameter
\begin{equation}
i\Delta_{DDW} Y({\bf k})= \langle c^{\dag}_{k+Q\sigma}c_{k\sigma}\rangle
\end{equation}
for a real constant $\Delta_{DDW}$.
Although the  $d$-density wave state appears as a solution of the mean-field 
equations, this state is hardly stabilized in variational Monte
Carlo calculations.  Thus, we are motivated to consider the lattice distortion,
which is going to stabilize the $d$-density wave cooperating with the spin-orbit
coupling. 

The inhomogeneous ground state under the lattice distortion in the underdoped
region has been investigated intensively using the two-dimensional (2D) 
Hubbard model.[13-16]
Within the 2D Hubbard model, the linear dependence of $\delta$ on $x$ has been
explained by variational Monte Carlo (VMC) methods[16], and furthermore, the
saturation of incommensurability $\delta$ for $x>0.125$ is also consistent with
those obtained by the VMC methods.
There is also a tendency towards the formation of stripes under the lattice 
distortions, with vertical or
horizontal hole-rich arrays coexisting with incommensurate magnetism and 
superconductivity (SC).[7]

We are going to evaluate spectral functions in the lightly to
optimally doped regions.
The tilt angle for LSCO is estimated as $14^{\circ}$-$18^{\circ}$ by 
EXAFS.[17]
Thus, $\theta$ is approximately $0.1\pi$.  If we use $\lambda\sim 0.2$,
we have $\phi=\lambda\theta\approx 0.02\pi$.
Here, we use a slightly larger value, $\phi=0.05\pi$, in actual calculations to
obtain
sufficient precision because of
the numerical difficulty for the small splitting.
Clearly, we can expect that this does not change the peculiar feature of the 
spectra.
We determine the variational parameters $g$ and $\Delta_Q$ so as to minimize
the ground state energy for $U=4$ and the carrier density in the range of
$0<x<0.2$ by the VMC methods.
The Hamiltonian is
\begin{equation}
H= H_0+U\sum_in_{i\uparrow}n_{i\downarrow},
\end{equation}
where $H_0$ is given in eq. (1).
The wave function is written in a Gutzwiller form:
\begin{equation}
\psi= P_G\psi_{MF}.
\end{equation}
$P_G$ is the Gutzwiller operator
\begin{equation}
P_G=\prod_i(1-(1-g)n_{i\uparrow}n_{i\downarrow}), 
\end{equation}
and
the mean field wave function $\psi_{MF}$ is obtained as an eigenfunction of
the Hartree-Fock Hamiltonian
\begin{equation}
H_{MF}= H_0+\sum_{i\sigma}[\delta n_i-{\rm sign}(\sigma)(-1)^{x_i+y_i}m_i]
c^{\dag}_{i\sigma}c_{i\sigma}.
\end{equation}
$\delta n_i$ and $m_i$ are expressed by the modulation vectors $Q_s$ and $Q_c$
for the spin and charge part, respectively.[7]
Equivalently, we use the form given by[7,18]
\begin{equation}
\delta n_i=-\alpha\sum_j{\rm cosh}((x_i-x_j^{str})/\xi_c), 
\end{equation}
and
\begin{equation}
m_i=\Delta_Q\prod_j{\rm tanh}((x_i-x_j^{str})/\xi_s).
\end{equation}
The diagonally striped state has hole arrays in the diagonal direction, while 
the
vertically striped state has hole arrays in the direction parallel to the $x$  
or $y$ direction.
Alternatively, it is also possible
to determine the order parameter
$\Delta_{\ell Q_s\sigma}=\sum_k\langle c^{\dag}_{k+\ell Q_s\sigma}c_{k\sigma}\rangle$
($\ell=0,1,\cdots,M-1$) for $\delta=1/M$ consistently.[19]
We obtain the diagonally striped state for $x<0.05$ and the vertically striped 
state for
$x>0.05$ as the ground state by the VMC methods.[16]
Among the diagonal stripes, the bond-centered striped state is the most stable 
when
we vary the wave function from the site-centered stripe to the bond-centered 
stripe by the VMC methods.
We mention here that the spin-orbit coupling stabilizes the diagonally striped
state in the light doping region.[20]

\begin{figure}
\includegraphics[width=9.5cm]{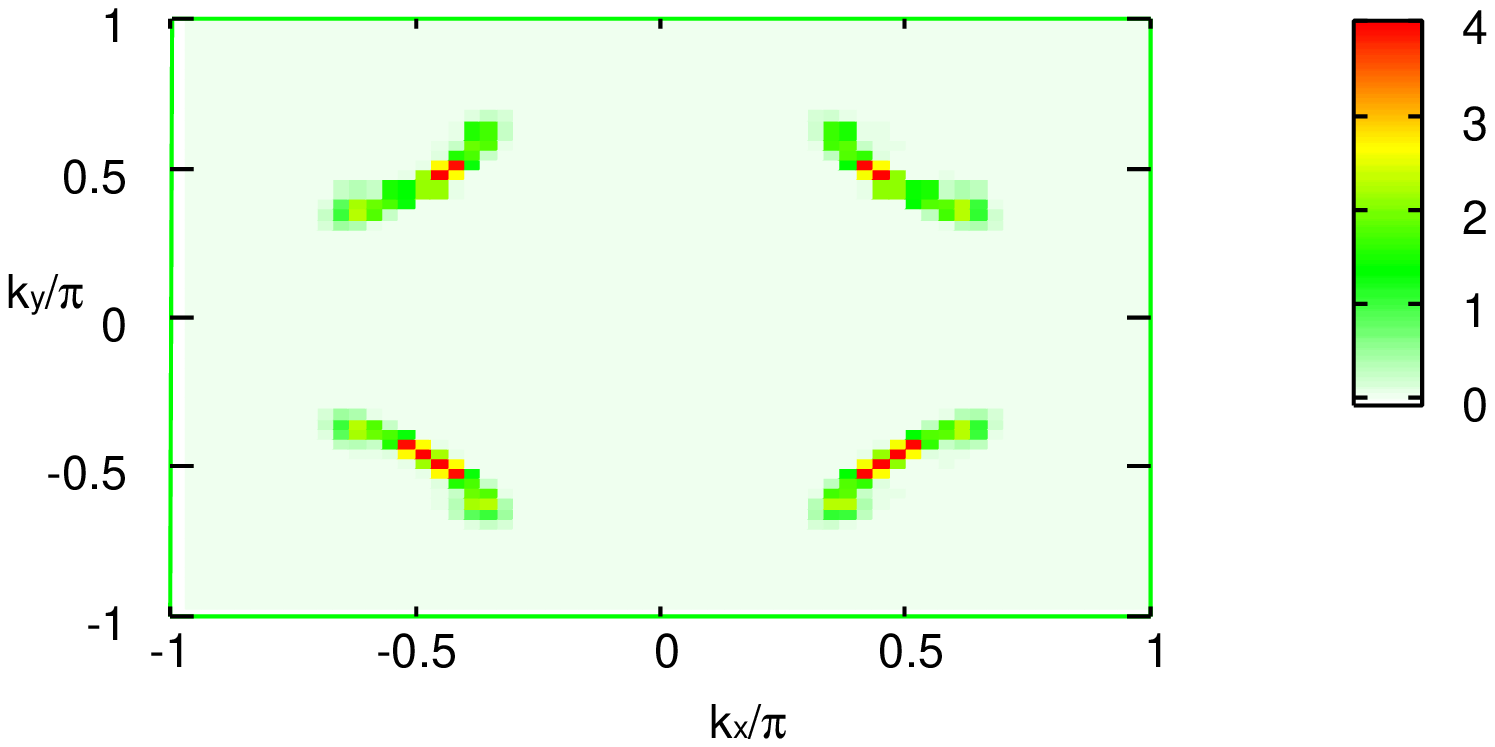}
\caption{
Contour map of density of states for the diagonally striped state 
at doping rate $x=0.03$
with $\phi/\pi=0.05$.
$\Delta_Q=0.08$, $\alpha=0$ and $t'=-0.2$.
}
\label{spec1}
\end{figure}

\begin{figure}
\includegraphics[width=9cm]{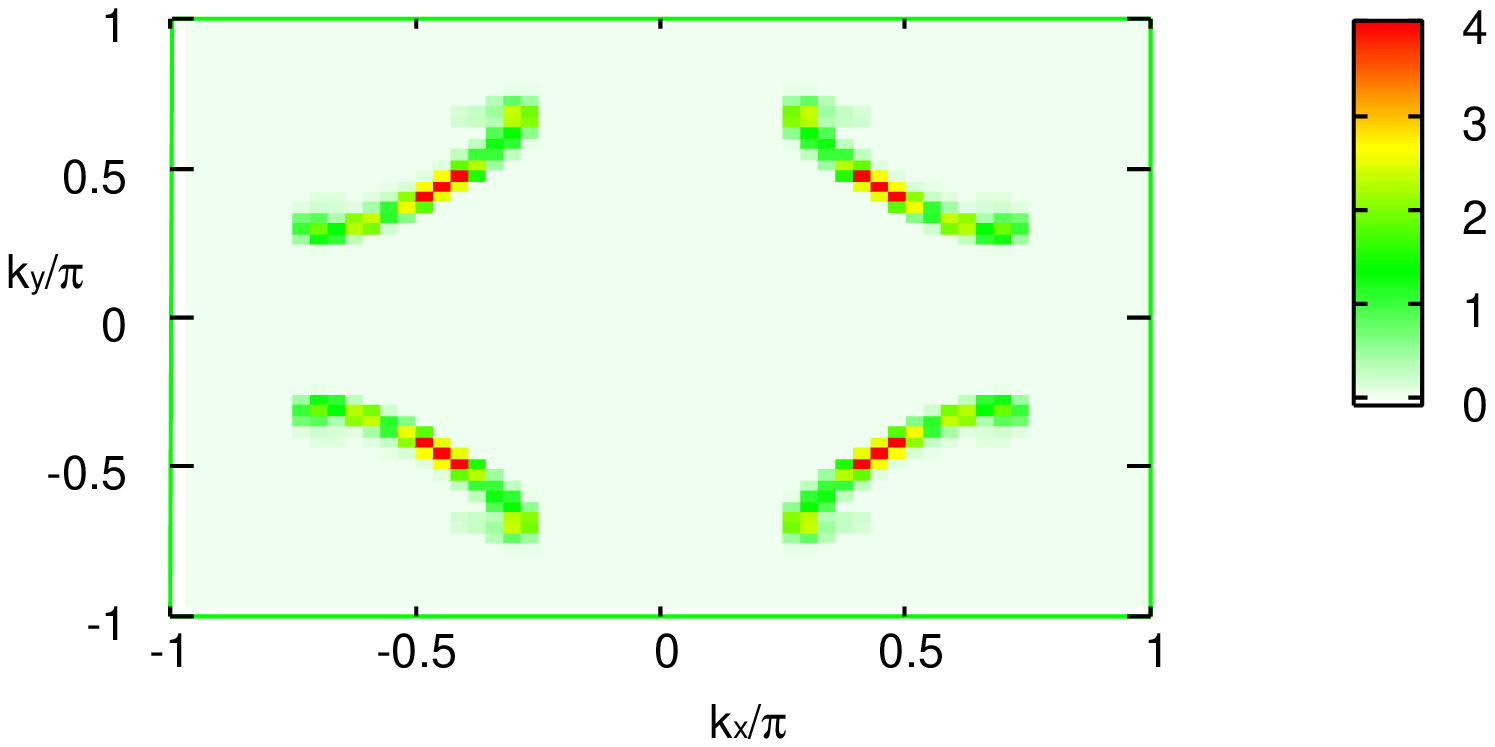}
\caption{
Density of states at doping rate $x=0.0612$ with $\phi/\pi=0.05$.
Vertical stripes with a 16-lattice periodicity are assumed.
$\Delta_Q=0.10$, $\alpha=0$ and $t'=-0.2$.
}
\label{spec2}
\end{figure}

\begin{figure}
\includegraphics[width=9cm]{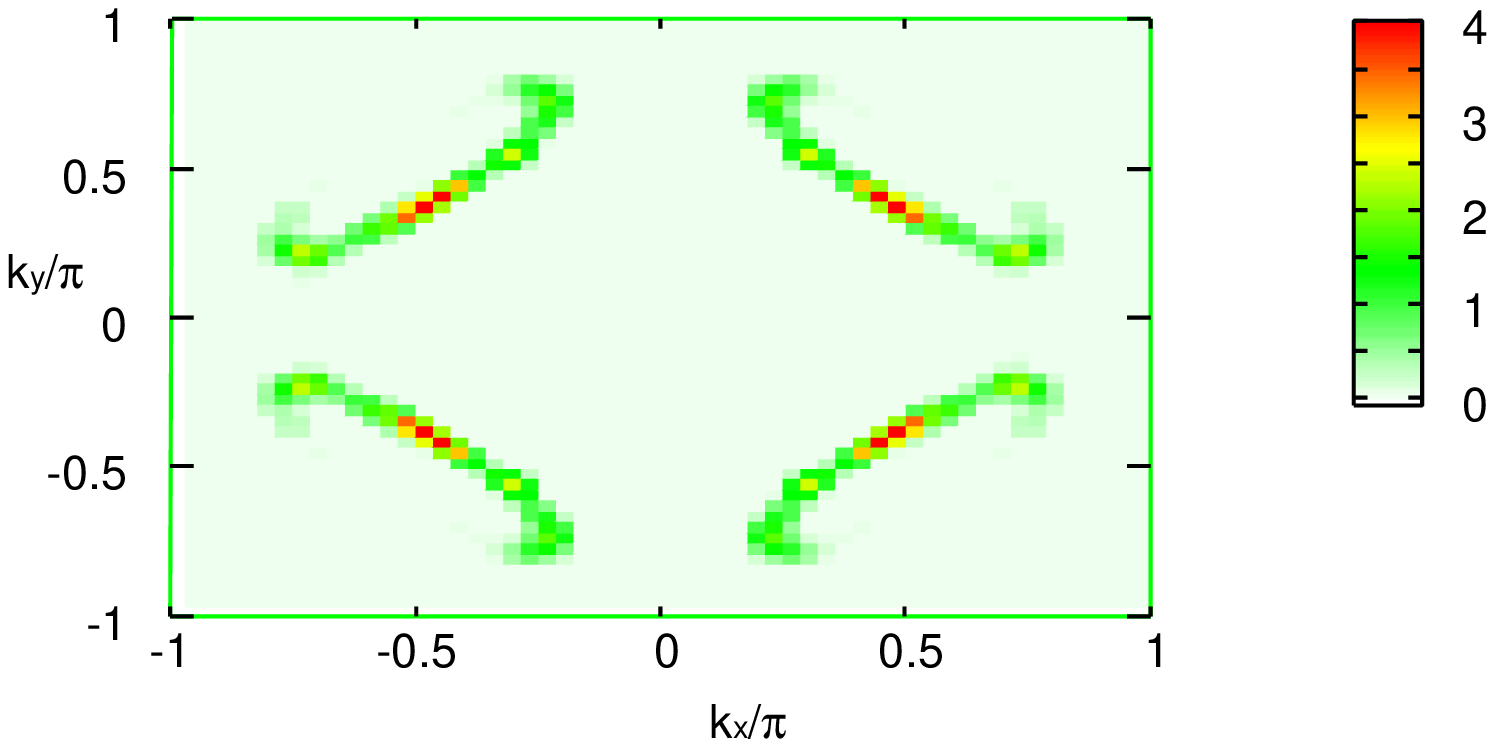}
\caption{
Density of states at doping rate $x=0.125$ with $\phi/\pi=0.05$.
Vertical stripes with an 8-lattice periodicity are assumed.
$\Delta_Q=0.16$, $\alpha=0$ and $t'=-0.2$.
}
\label{spec3}
\end{figure}

\begin{figure}
\includegraphics[width=9cm]{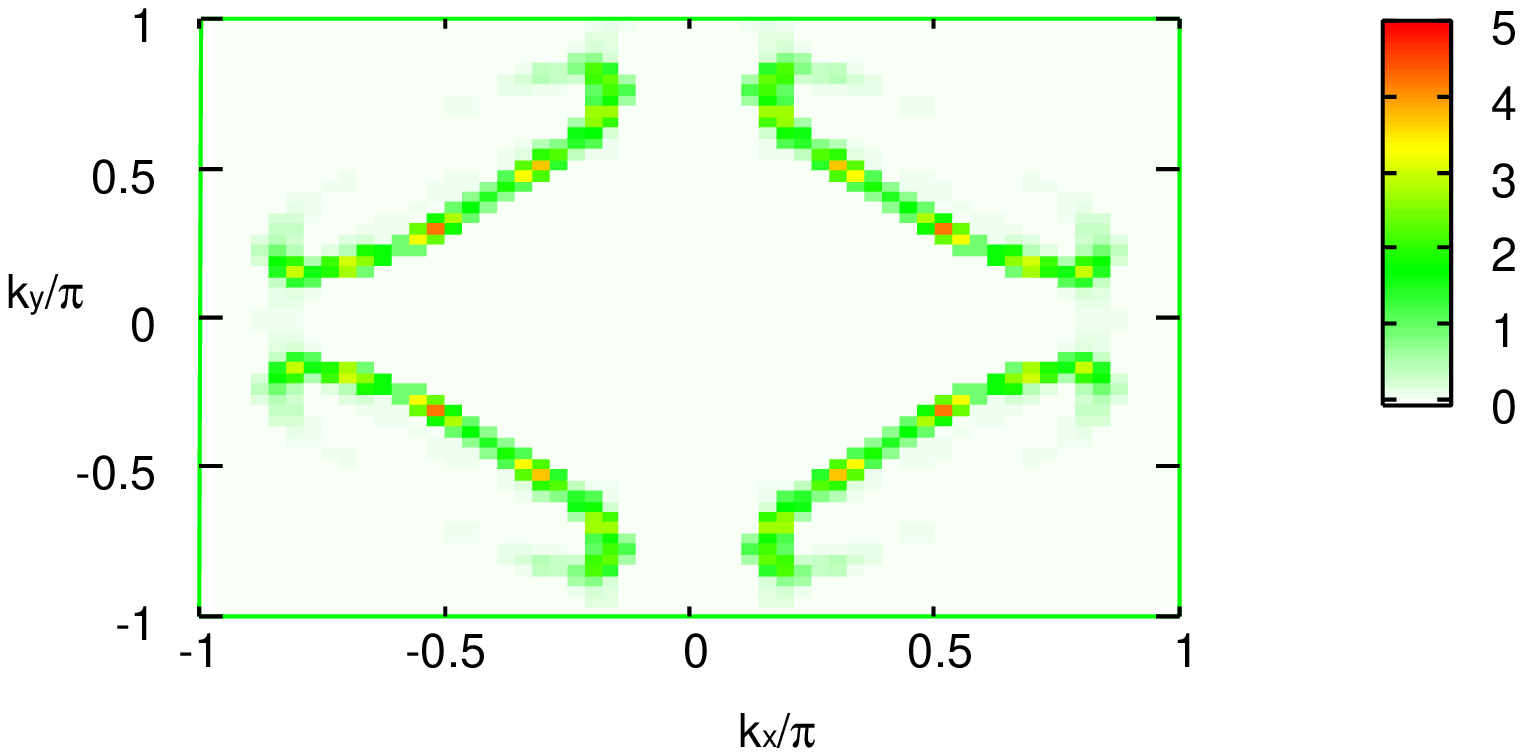}
\caption{
Density of states for the vertically striped state at doping rate $x=0.197$
with $\phi/\pi=0.05$.
Vertical stripes with an 8-lattice periodicity are assumed.
$\Delta_Q=0.08$, $\alpha=0$ and $t'=-0.2$.
}
\label{spec4}
\end{figure}

\begin{figure}
\includegraphics[width=9cm]{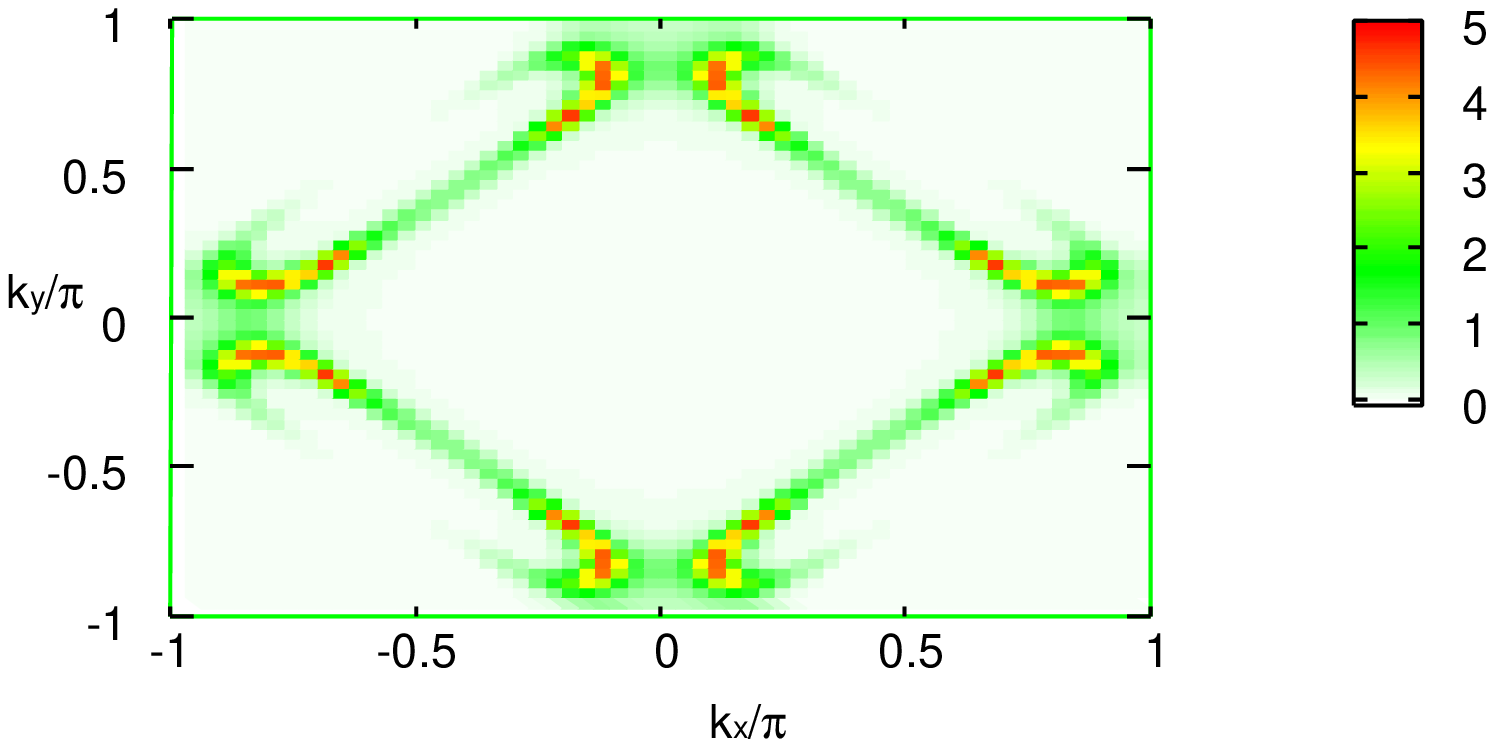}
\caption{
Density of states for the normal state at doping rate $x=0.197$
with $\phi/\pi=0.05$, $\Delta_Q=0$ and $t'=-0.2$.
}
\label{spec5}
\end{figure}

In the evaluations of spectral functions, we consider the effect of $P_G$
within the mean field theory since
we must consider the excited states as well as the ground state.
We then evaluate the spectral weight from the eigenvalues $E_{\sigma m}$ and 
eigenfunctions
$(u_{\sigma m})_j$ of the Hamiltonian $(H^{\sigma}_{MF})_{ij}$ in the real-space
representation, where $i$ and $j$ are site indices.  
Green's functions are
\begin{equation}
g_{\sigma}(i,j,i\omega)= \sum_m \frac{(u_{\sigma m})_i(u_{\sigma m})_j^*}
{i\omega-E_{\sigma m}},
\end{equation}
\begin{equation}
g_{\sigma}({\bf k},i\omega)=\frac{1}{N_a}\sum_{ij}{\rm e}^{-i{\bf k}\cdot({\bf R}_i
-{\bf R}_j)} g_{\sigma}(i,j,i\omega),
\end{equation}
where $N_a$ is the number of atoms.
The spectral function is calculated using the formula
\begin{equation}
N_{\sigma}({\bf k},\epsilon)= -\frac{1}{\pi}{\rm Im}g_{\sigma}({\bf k},\epsilon
+i\delta).
\end{equation}

We show the results for $x=0.03$, 0.061, 0.125 and 0.197 for the parameters 
obtained by
the VMC methods.
The contour map of spectra for the light-doping $x=0.03$ is shown in 
Fig. \ref{spec1}
where
the calculations were performed on a $60\times 60$ lattice for the half-filled
diagonal stripes.
A peak near $(\pi/2,\pi/2)$ appears due to the spin-orbit and distortion effects
as presented in Fig. \ref{spec1} for $\phi/\pi=0.05$, while it should be noted 
that the spectra of diagonal stripes without distortion exhibit a
one-dimensional structure in the diagonal direction.
It has been pointed out that the diagonally striped state can alone explain 
the opening
of the gap if we consider the bond-centered stripe.[21]
The spectral functions for $x=0.061$ and $x=0.125$ with $\phi/\pi=0.05$ are 
shown
in Figs. \ref{spec2} and \ref{spec3}, respectively, where we have vertically
striped states that have 16-lattice and 8-lattice periodicities,
respectively, in accordance with neutron scattering measurements.[4]
In Fig. \ref{spec4} the spectral map at $x=0.197$ in the overdoped region is 
shown,
where the stripes have an 8-lattice periodicity.
We show the spectra without stripes at $x=0.197$ in Fig. \ref{spec5} for 
comparison.
We observe the absence of spectral weight near $(\pi,0)$, which is a 
characteristic
structure originating from the pseudogap.
The vertically striped state has one-dimensional-like spectra near 
($\pm\pi$,0) with the
Fermi wave number $k_F$ corresponding to the one-dimensional quarter-filled
band,[13]
while the $\Delta_k$ term contributes to a peak structure at around
$(\pm\pi/2,\pm\pi/2)$.
As a result, we obtain the arclike spectra for vertical stripes.
Thus, as the doping rate $x$ increases, the spectra near $(\pi/2,\pi/2)$ in the
light-doping case extends towards the 2D-like Fermi surface in the optimally 
doped region, which occurs as a crossover.

In this paper we have examined novel phenomena stemming from the spin-orbit 
coupling
induced by the tilting of CuO octahedra. 
We have shown that the characteristics of the spectral function of doped 
high-$T_c$
cuprates can be consistently explained using the 2D electronic model with 
lattice
distortions, which is in contrast to a theory considering $p$-orbitals in 
apical oxygen atoms. [22] 
We have taken into account the incommensurate structure observed by
neutron-scattering experiments.
Furthermore, we can expect the following fascinating
physics: pseudogap, nodal metal, time-reversal symmetry breaking,
diagonal stripes, and string-density wave as a generalization of the
$d$-density wave.
In the half-filled case, we obtain a Fermi
point, and thus, a peak exists near $(\pi/2,\pi/2)$ in the density of states
that extends to form the Fermi arc spectra as the doping rate 
increases.[2]
In the low-doping case, we obtain a nodal metallic state in the diagonally 
striped state, 
since the pseudogap has a $d_{x^2-y^2}$ symmetry, which is consistent with the
experiments of resistivity.[6]
The arc is expected to expand with increasing temperature to form the full
Fermi surface above the splitting gap $\sim 100$K.[1]
Lastly, we comment on a possibility of superconductivity along stripes.
The coexistence of superconductivity and stripes has been pointed out for 
vertical
stripes by the VMC methods.[7]  It is difficult to have a stable coexistent 
state of
SC and stripes for diagonal stripes in the VMC methods because the SC pairs 
must have a $d_{xy}$ 
symmetry along stripes.
Thus, superconductivity is suppressed for light doping.

We are grateful to H. Eisaki for valuable discussions.




\end{document}